\documentclass[3p,times]{elsarticle}
\usepackage{ecrc}
\usepackage{hyperref}

\usepackage{epstopdf,color}

\volume{00}

\firstpage{1}

\journalname{Nuclear Physics A}

\runauth{Yan et al.}

\jid{nupha}

\jnltitlelogo{Nuclear Physics A}

\usepackage{graphicx}
\usepackage{amsmath,amssymb}
\def\be{\begin{eqnarray}}
\def\ee{\end{eqnarray}}

\def\Eq#1{Eq.~(\ref{#1})}
\def\Eqs#1{Eqs.~(\ref{#1})}

\def\Fig#1{Fig.~\ref{#1}}
\def\Sect#1{Section~\ref{#1}}
\def\Ref#1{~\cite{#1}}

\def\etc{{\it etc.}}

\begin{document}

\begin{frontmatter}



\title{Universal parameterization of initial-state fluctuations and its applications to event-by-event 
anisotropy}

\author[lyjo]{Li Yan} 
\author[lyjo]{Jean-Yves Ollitrault}
\author[ap]{Art M. Poskanzer}

\address[lyjo]{CNRS, URA2306, IPhT, Institut de Physique Th\'eorique de Saclay, F-91191 Gif-sur-Yvette, France}
\address[ap]{Lawrence Berkeley National Laboratory, Berkeley, California, 94720}

%% For multiple authors, replace the above by:

%\author[label1]{Author1}
%\author[label2]{Author2}

%\address[label1]{Address 1}
%\address[label2]{Address 2}

\begin{abstract}
We propose Elliptic Power and Power parameterizations for the probability distribution 
of initial state anisotropies in heavy-ion collisions.
%proton-proton, proton-nucleus and 
%nucleus-nucleus collisions.
%Accounting for the only fact that these anisotropies are bounded by 
%unity, we find that an 'Elliptic Gaussian' distribution is sufficient in 
%describing pure
%fluctuation-driven eccentricities, such as triangularity $\epsilon_3$ of 
%the initial state in lead-lead collisions; and eccentricities with 
%non-zero expectations, such as
%ellipticiticy $\epsilon_2$ of the initial state in deuteron-gold 
%collisions. Similar tests have been carried out in various Monte-Carlo 
%simulations, irrespective of the detailed
%construction of the collision models. 
By assuming a linear eccentricity scaling, the new parameterizations can also be applied to
fluctuations of harmonic flow. In particular, 
we analyze flow multi-particle cumulants and event-by-event distributions, both of which
are recently measured at the LHC.
%the recently measured flow multi-particle cumulants and  
%event-by-event distributions are analyzed. 
%In addition to the 
%consistent characterization of event-by-event distributions of elliptic 
%flow $v_2$, and triangular
%flow $v_3$ from ATLAS collaboration with the rescaled 
%parameterizations, the flow response coefficient of these harmonics 
%are extracted, independently
%of any detailed information of the initial state.
\end{abstract}

\begin{keyword}

%Keyword1 \sep Keyword2 \sep Keyword3
Heavy-ion collisions \sep fluctuations \sep anisotropic flow

\end{keyword}

\end{frontmatter}

% \linenumbers


\section{Introduction}
\label{intro}

It was recently realized that the understanding of fluctuations, in particular fluctuations in the initial state, 
is an essential ingredient in the analyses of ultra-relativistic heavy-ion collisions\Ref{Alver:2010gr}. 
To characterize a fluctuating initial state %and its effects on experimental observables, 
theoretically, effective models have been proposed by properly introducing fluctuations  
on top of nucleus-nucleus collisions\Ref{Alver:2008aq,Schenke:2014tga}. However, despite some success of these models, 
the initial state of heavy-ion collisions still contributes a major fraction of the uncertainty 
of quantitative predictions\Ref{Retinskaya:2013gca,Luzum:2012wu}.
In experiments, initial state fluctuations can be revealed by the study of anisotropic 
flow $v_n$. Defined as the Fourier harmonics of the corresponding particle spectrum, $v_n$ reflects the property of bulk 
medium expansion, and its response to the initial state anisotropy. Taking into account thus the direct mapping between $v_n$ and initial 
anisotropy, which is commonly formulated as eccentricity $\varepsilon_n$, it is expected that event-by-event (EbyE) distribution of
$v_n$ is largely determined by fluctuations of $\varepsilon_n$. Indeed, many non-trivial observations have been 
made in heavy-ion experiments regarding flow fluctuations, 
among which the EbyE distribution of $v_n$ in Pb-Pb collisions\Ref{Aad:2013xma}, and cumulants of elliptic 
flow $v_2$ from p-Pb collisions\Ref{Chatrchyan:2013nka,CMS:2014bza} will be discussed in this work. 
In this paper, without detailed modeling of initial state we propose a new parameterization
to describe $\varepsilon_n$ fluctuations. As will be shown in \Sect{modeling}, the crucial improvement of our new parameterization is rooted in the fact that 
$|\varepsilon_n|\le 1$. The universality of parameterizing fluctuations of $\varepsilon_n$ will be addressed also in \Sect{modeling}.  
In \Sect{app} we apply the parameterization to the measured flow cumulants and flow distribution.  
%Conclusions will be made in \Sect{sum}.

\section{Elliptic-Power and Power parameterizations}
\label{modeling}

Initial state eccentricity characterizes the spatial anisotropy of a system created in heavy-ion collisions. 
Ellipticity, for example, which can be defined in a complex form as 
$\varepsilon_2 e^{i2\Phi_2}=\varepsilon_{2x}+i\varepsilon_{2y}=-\{r^2 e^{i2\phi_r}\}/\{r^2\}$,
characterizes the elliptic deformation. 
In the definition, $\{\ldots\}=\int dxdy \ldots\epsilon(x,y)$ stands for an average in the transverse plane with respect to energy density
$\epsilon(x,y)$,
which implies that the modeling of eccentricity replies on an effective description of energy deposition from 
nucleon-nucleon collisions. In addition, fluctuations and correlations 
in the colliding system need to be 
included as well, so that initial state eccentricity fluctuates on an EbyE basis. 
Nonetheless, taking into account fluctuations of nucleons as a dominant effect 
one arrives at the ``independent source model". Nucleon-nucleon 
correlations are ignored in such a simplified description, but effects of fluctuation can be solved analytically. Following
similar procedures taken in the original applications\Ref{Ollitrault:1992bk} 
and further assuming, such as Gaussian background \etc,
for a system configured by $N$ independent 
point-like sources, the spatial anisotropy is found to fluctuate according to the Elliptic-Power distribution\Ref{Yan:2014afa},
\be
\label{EPdis}
P(\varepsilon_x,\varepsilon_y)=\frac{\alpha}{\pi}(1-\varepsilon_0^2)^{\alpha+\frac{1}{2}}
\frac{(1-\varepsilon_x^2-\varepsilon_y^2)^{\alpha-1}}{(1-\varepsilon_0\varepsilon_x)^{2\alpha-1}}\,.
\ee 
\Eq{EPdis} contains two parameters. $\alpha\sim N$ is approximately determined by the magnitude of fluctuations.
$\varepsilon_0\sim \varepsilon_{RP}$ is constrained by the event-averaged eccentricity,
where $\varepsilon_{RP}$ is the generally defined reaction-plane ellipticity. 
For the cases when event-averaged energy density is azimuthally symmetric,
such as proton-lead collisions carried out recently at the LHC\Ref{Chatrchyan:2013nka,CMS:2014bza}, $\varepsilon_0=0$. Then \Eq{EPdis} reduces to 
the Power distribution\Ref{Yan:2013laa}
\be
\label{Power}
P(\varepsilon_x,\varepsilon_y)=\frac{\alpha}{\pi}(1-\varepsilon_x^2-\varepsilon_y^2)^{\alpha-1}\,.
\ee

\begin{figure}
\begin{center}
\includegraphics[width=0.4\textwidth] {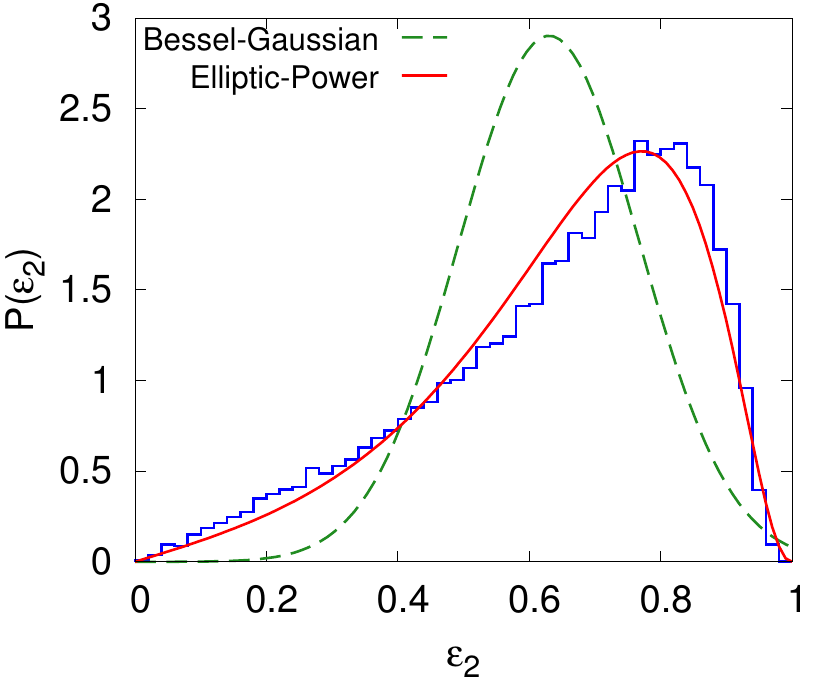}
\includegraphics[width=0.4\textwidth] {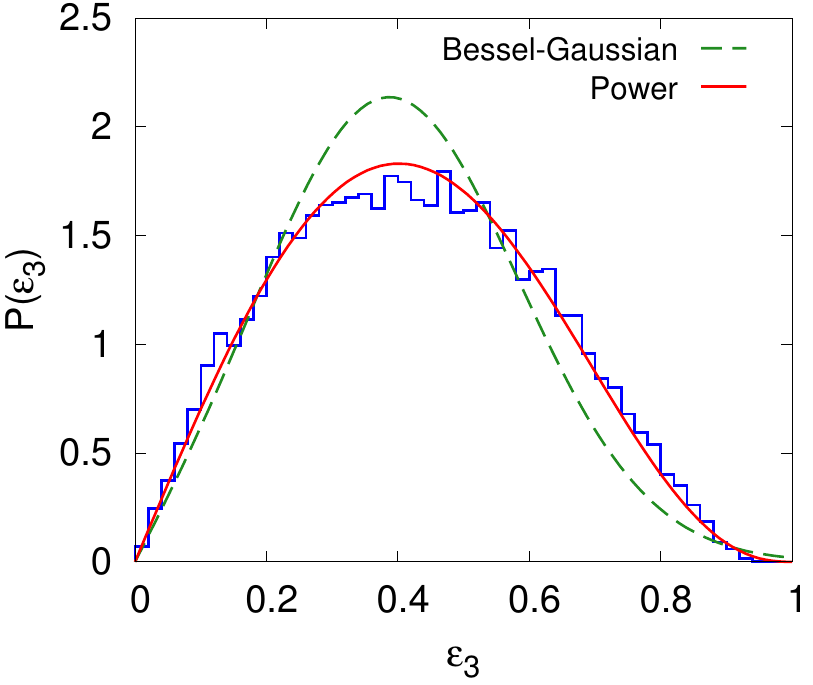}
\caption{
\label{ep2ep3}
(Color online) Distribution of $\varepsilon_2$ and $\varepsilon_3$ in 75-80\% central Pb-Pb collisions.
Histograms are obtained by PHOBOS Monte-Carlo Glauber simulations. Fit by Bessel-Gaussian
parameterization (green dashed lines) are shown comparing to Elliptic Power (red solid line in the
left panel)
for $\varepsilon_2$ and Power (red solid line in the right panel) for $\varepsilon_3$. 
}
\end{center}
\end{figure}

Integrating out the angular dependence in \Eqs{EPdis} and (\ref{Power}), we obtain 1-dimensional Elliptic-Power and Power 
parameterizations for EbyE fluctuating eccentricities. Note that the integration of angle in \Eq{EPdis} results in a 
hyper-geometric function which can be done in practice numerically. We have tested both of these new parameterizations by
comparing to the eccentricity distributions obtained by Monte-Carlo simulations of Glauber\Ref{Alver:2008aq} and 
IP-Glasma\Ref{Schenke:2014tga} models, and reasonably good fits are found universally in all
collision centralities\Ref{Yan:2014afa}. In \Fig{ep2ep3} we present one of such comparisons for collisions with 
centrality percentile $75-80\%$. % for PHOBOS Monte-Carlo Glauber model. 
We notice that our new parameterizations respect the fact that $\varepsilon_n$ is bounded by unity. This crucial
property leads to a significant improvement compared to the Bessel-Gaussian distribution, 
%which has been taken as the ``state-of-art" description of eccentricity fluctuations.
especially for small systems where the $\varepsilon_n$ values are larger.
% and been applied in the corresponding experiments\Ref{Aad:2013xma}.

\section{Applications to p-Pb and Pb-Pb systems}
\label{app}

\begin{figure}
\begin{center}
\includegraphics[width=0.4\textwidth] {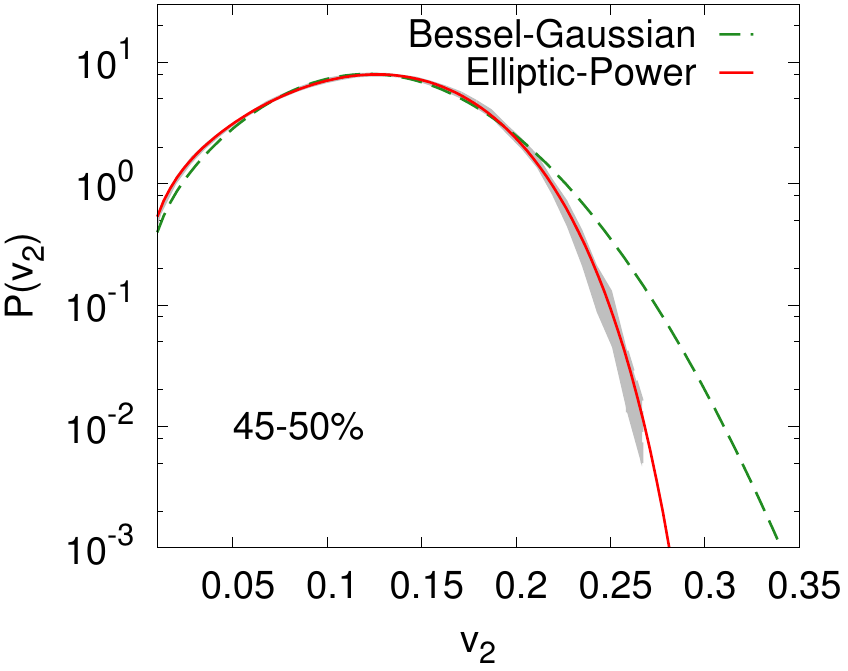}
\includegraphics[width=0.378\textwidth] {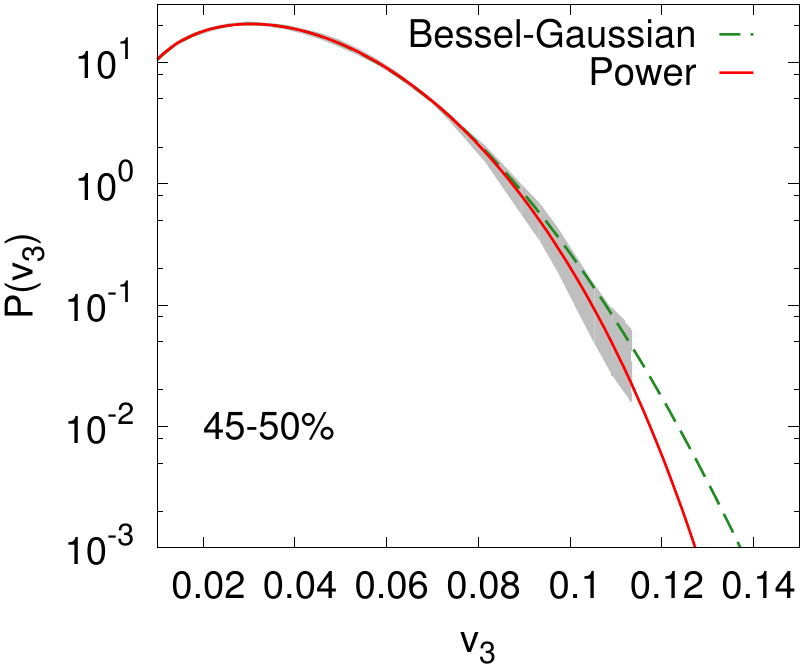}
\caption{
\label{v2v3}
(Color online) Fits of ATLAS\Ref{Aad:2013xma} flow event distribution (grey band) from $45-50\%$ central Pb-Pb collisions, 
with Elliptic Power and Power distributions (red solid lines). Bessel-Gaussian fits (green dashed lines)
are shown for comparison. 
}
\end{center}
\end{figure}

Medium response to the initial state of heavy-ion collisions converts spatial anisotropy into anisotropic flow. In particular,
for elliptic flow $v_2$ and triangular flow $v_3$, it is known that the response is to a good approximation linear (see for 
instance\Ref{Niemi:2012aj}), namely, 
$
v_2e^{i2\Psi_2}=\kappa_2\varepsilon_2e^{i2\Phi_2} $
and $
v_3e^{i3\Psi_3}=\kappa_3\varepsilon_3e^{i3\Phi_3}\,. 
$
The linear eccentricity scaling allows us to apply \Eqs{EPdis} and (\ref{Power}) to flow fluctuations, after a rescaling with the flow response coefficients $\kappa_2$ or $\kappa_3$. 

%To analyze flow fluctuations, one way is to
One application is the study of multi-particle cumulants of the measured flow. It should be emphasized that 
%as being constructed in a specified form, 
flow cumulants have non-trivial dependence on the nature of fluctuations. 
%For example, second order
%cumulant $v\{2\}=\sqrt{\bra v^2\ket}$ measures the mean value, while fourth order cumulant $v\{4\}=(2\bra v^2\ket^2-\bra v^4\ket)^{1/4}$
%roughly determines the variance. 
For example, one may check that for a pure Gaussian fluctuation, higher order flow cumulants
vanish by definition. For the Power distribution in \Eq{Power}, we notice that cumulants of any order can be analytically
expressed as a function of $\alpha$\Ref{Yan:2013laa}. In addition, it has also been found that higher order cumulants of the Power 
distribution follow a certain pattern, which results in determined relations among cumulant ratios. Recent measurements by the CMS
collaboration\Ref{Chatrchyan:2013nka,CMS:2014bza} confirms that $(v_2\{4\}/v_2\{2\})$ and $(v_2\{6\}/v_2\{4\})$ from proton-lead collisions are 
quantitatively consistent with Power distribution predictions, which strongly supports the picture of collective expansion
in the p-Pb system\Ref{Yan:2013laa}.   

%More straightforwardly, one can analyze flow fluctuations by describing event-by-event flow distribution, as measured by ATLAS collaboration
%for Pb-Pb collisions\Ref{Aad:2013xma}. Taking into account \Eq{EPdis} and \Eq{lresp}, elliptic flow is expected to fluctuate according
%to the rescaled Elliptic-Power distribution, as long as one replaces $\varepsilon_2$ everywhere in \Eq{EPdis} by $v_2/\kappa_2$. One thus
%obtain a new parameterization as function of $\alpha$, $\varepsilon_0$ and $\kappa_2$, which can be applied to the experimentally 
%measured $v_2$ distribution using, for example, least-$\chi^2$ fit.  

\begin{figure}
\begin{center}
\includegraphics[width=0.33\textwidth] {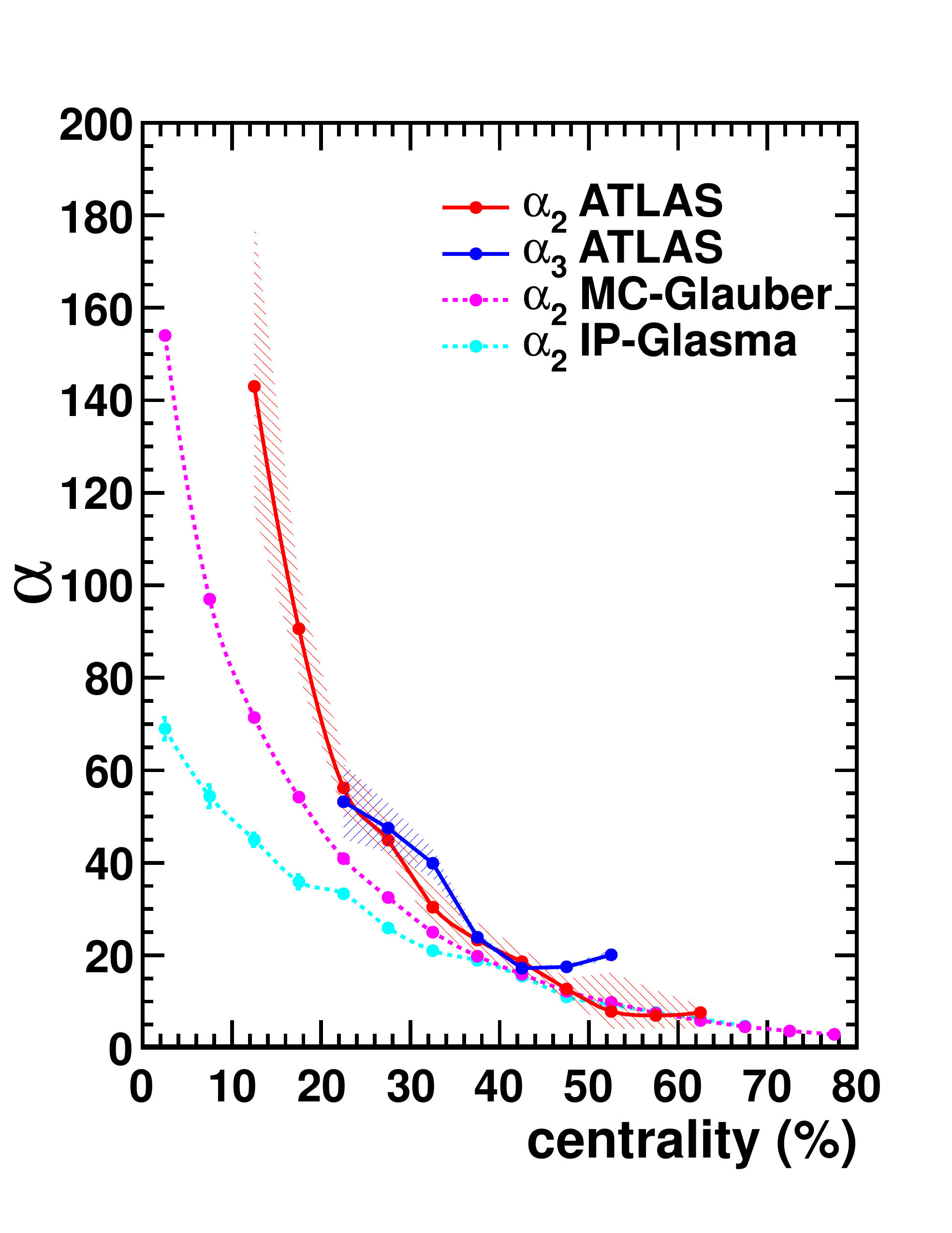}
\includegraphics[width=0.33\textwidth] {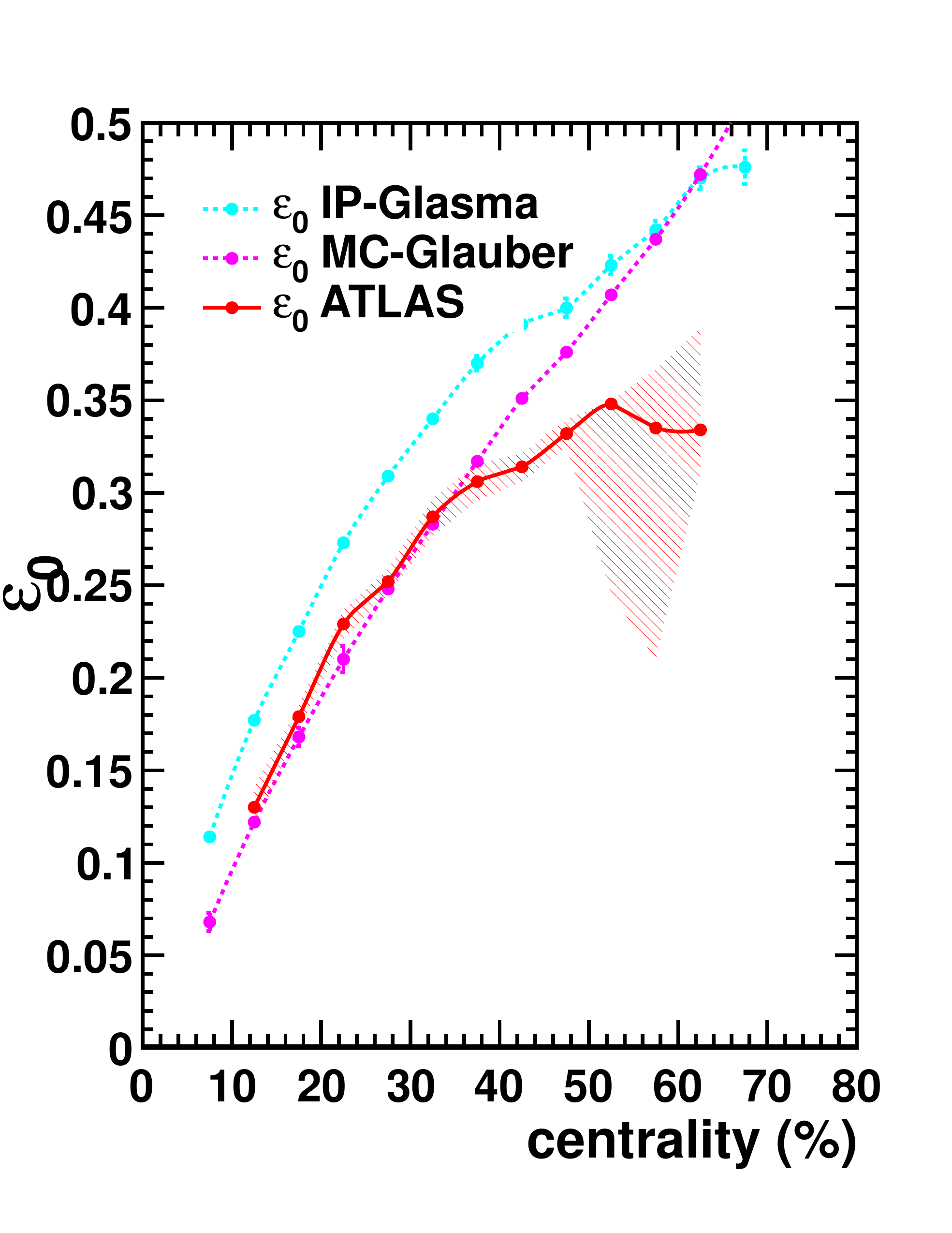}
\includegraphics[width=0.33\textwidth] {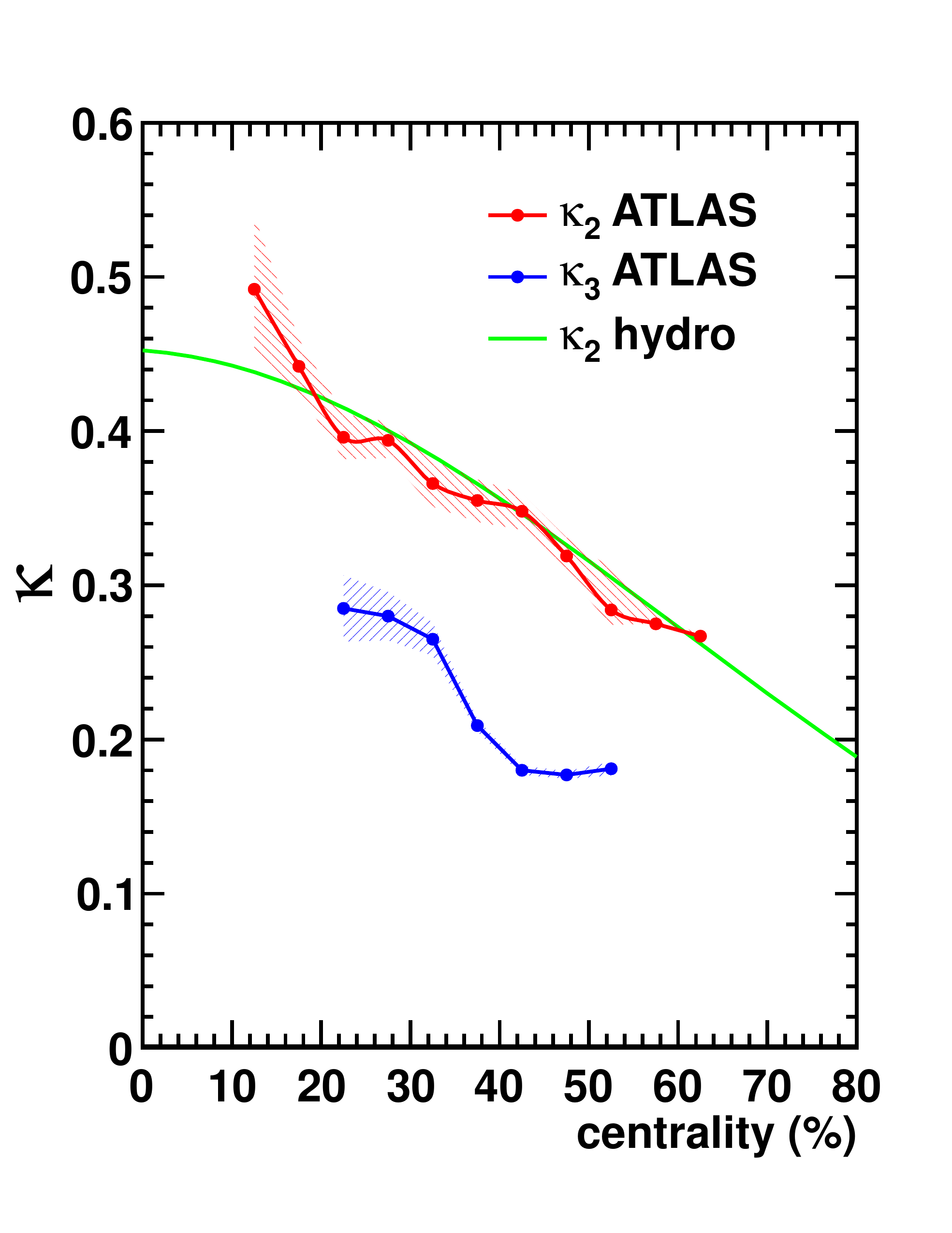}
\caption{
\label{fitpars}
(Color online) Parameters extracted from fit of ATLAS $v_2$ event distribution. $\alpha$ and $\varepsilon_0$ from Glauber and IP-Glasma models
are present for comparisons. $\kappa_2$ from hydro predictions with $\eta/s\sim0.18$ is shown as green solid line in the right panel.
}
\end{center}
\end{figure}

Alternatively flow fluctuations can be analyzed by fitting EbyE flow distribution 
with Elliptic-Power or Power parameterizations. Replacing $\varepsilon_2$ everywhere in \Eq{EPdis} by $v_2/\kappa_2$, one finds a 
rescaled Elliptic-Power distribution, as a function of $\alpha$, $\varepsilon_0$ and $\kappa_2$. In a similar manner, for $v_3$ the
substitution $\varepsilon_3=v_3/\kappa_3$ leads to rescaled Power distribution as a function of $\alpha$ and $\kappa_3$. In \Fig{v2v3},
we present fits of ATLAS measured EbyE $v_2$ and $v_3$ distribution from $45-50\%$ centrality Pb-Pb collisions\Ref{Aad:2013xma}, 
with Elliptic-Power and Power parameterizations. 
Comparing to a Bessel-Gaussian, it is clear that both Elliptic-Power and Power distributions achieve better agreement with experiments.
Similar fits can be extended to all centrality bins, and as expected, we found that 
the improvements with Elliptic Power and Power parameterizations are more pronounced as centrality percentile grows. 
In \Fig{fitpars} the extracted parameters from the fitting procedure are shown as a function of centrality. Shaded area of $\alpha_2$,
$\epsilon_0$ and $\kappa_2$ are associated with systematic and statistical errors of the measured $v_2$, 
while for $\alpha_3$
and $\kappa_3$ only the effect of statistical errors is considered.
We found that systematic errors of $v_3$ lead to anomalously large uncertainties of $\alpha_3$ and $\kappa_3$, 
which makes the results from ATLAS $v_3$ less meaningful. Nevertheless, 
we leave $\alpha_3$ and $\kappa_3$ in \Fig{fitpars}. It should be noticed that the way of obtaining these parameters, 
especially flow response coefficients $\kappa_2$ and $\kappa_3$, relies very little on the detailed modeling of initial states.
The left panel of \Fig{fitpars} shows a decrease of 
$\alpha$ with respect to centrality percentile, which indicates an increase of initial state fluctuations from central to 
peripheral collisions. Also, we find that the average geometry is more elliptic towards 
peripheral collisions, as being depicted by the growth of $\epsilon_0$ in the middle of \Fig{fitpars}.
Both of these results are consistent with our na\"ive understanding. We also plot in \Fig{fitpars} the corresponding 
predictions by PHOBOS MC-Glauber and IP-Glasma models. 
Flow response coefficient
$\kappa_2$ is solely determined by the bulk property of the medium. As shown in the right panel of \Fig{fitpars}, $\kappa_2$ has a 
clear trend of decreasing when the system is getting smaller, as anticipated by hydrodynamics.
Using a 2+1D viscous hydrodynamics (details of our hydrodynamics modeling can be found, for instance, in\Ref{Teaney:2013gca}), 
we found that $\kappa_2$ is well described as a function of shear viscosity over entropy ratio $\eta/s$,
\be
\label{kappa_hydro}
\kappa_2(\eta/s)=C_0\left[\kappa_2^{\mbox{\tiny ideal}}-\frac{\eta}{s}\delta \kappa_2\right]\,,
\ee
where $\delta \kappa_2=-[\kappa_2^{\mbox{\tiny ideal}}-\kappa_2^{\mbox{\tiny visc.}}]/(1/4\pi)$ characterizes the change of response 
coefficients due to shear viscosity. $\kappa_2^{\mbox{\tiny ideal}}$ and $\kappa_2^{\mbox{\tiny visc.}}$  
are obtained via hydrodynamics, with $\eta/s$ specified to be zero and $1/4\pi$ respectively in simulations. Constant $C_0\sim 1.68$ in
\Eq{kappa_hydro} takes into account all the extra effect that is not included in our hydrodynamic calculations. 
\Eq{kappa_hydro} returns an estimate of $\eta/s\sim 0.18$.

\section{Summary and conclusions}
\label{sum}

We have proposed Elliptic Power and Power parameterizations for the initial eccentricity fluctuations. 
The 
validity of parameterizations are examined by fitting to effective models, 
with satisfactory agreements
universally found. We further apply rescaled Elliptic Power and Power parameterizations to the distribution of anisotropic flow $v_n$, 
using the linear eccentricity scaling. Fitting to experimental data allows us to extract parameters which are relevant to the information 
of the initial state, such as $\alpha$ and $\epsilon_0$. Also, and more importantly, the flow response coefficient $\kappa_2$ is obtained quantitatively 
without any effective modeling of the initial state, which provides a more self-contained way to estimate of the bulk property of the quark-gluon plasma.

\section*{Acknowledgments}

LY and JYO are funded  by the European Research Council under the
Advanced Investigator Grant ERC-AD-267258. 
AMP was supported by the Director, Office of Science, of the U.S. Department of Energy.



%% References


\bibliographystyle{iopart-num}
%\bibliographystyle{elsarticle-num}
%\bibliographystyle{utphys}
\bibliography{refs-qm14}



%\begin{thebibliography}{00}


%\bibitem{ref1} J. van der Geer, J.A.J. Hanraads, R.A. Lupton, J. Sci. Commun. 163 (2000) 51Ð59. 
%\bibitem{ref2} W. Strunk Jr., E.B. White, The Elements of Style, third ed., Macmillan, New York, 1979. 

%\end{thebibliography}

\end{document}